\documentclass[twocolumn,english,prb]{revtex4-1}
\usepackage[latin9]{inputenc}
\usepackage{float}
\usepackage{amsmath}
\usepackage{graphicx}
\usepackage{esint}
\usepackage{hyperref}
\hypersetup{
     colorlinks = true,
     citecolor  = red,  
     linkcolor  = magenta 
}
\usepackage[titletoc]{appendix}
\usepackage{bm}\usepackage{xcolor}
\usepackage{babel}
\makeatletter
\newcommand{\bise}{Bi$_2$Se$_3$ }

\newcommand{\smb}{SmB$_6$ }

\begin{document}

\title{Pseudo-Spin, Real-Spin and Spin Polarization of Photo-emitted Electrons}

\author{Rui Yu$^{1}$, Hongming Weng$^{2,3}$, Zhong Fang$^{2,3}$ Xi Dai$^{2,3}$}
\email{daix@iphy.ac.cn}

\affiliation{$^{1}$
Department of Physics, Harbin Institute of Technology, Harbin 150001, China}

\affiliation{$^{2}$
Beijing National Laboratory for Condensed Matter Physics, and Institute of Physics, Chinese Academy of Sciences, Beijing 100190, China}

\affiliation{$^{3}$
Collaborative Innovation Center of Quantum Matter, Beijing 100190, China}

\begin{abstract}
In this work, we discuss the connections between 
pseudo spin, real spin of electrons in material 
and spin polarization of photo-emitted electrons out of material.
By investigating these three spin textures 
for \bise and \smb compounds,
we find that the spin orientation of photo-electrons 
for \smb has different correspondence to pseudo spin and real spin 
compare to Bi$_2$Se$_3$, 
due to the different symmetry properties of the photo-emission
matrix between initial and final states.
We calculate the spin polarization and circular dichroism spectra
of photo-emitted electrons for both compounds, 
which can be detected by spin-resolved and
circular dichroism angle resolved photo-emission spectroscopy experiment.
\end{abstract}

\date{\today}

\pacs{xx}

\maketitle

\section{Introduction}

The experimental technique of angle resolved photo-emission spectroscopy 
(ARPES) is a powerful tool in investigating the electronic structure of
crystalline materials.
The energy and momentum information of the electrons inside a materials
can be obtained by measuring the kinetic energy and angular distribution
of the photo-emitted electrons from a sample illuminated with sufficiently
high-energy radiation.
To detect the spin information of the electronic states, the spin-resolved
ARPES and circular dichroism (CD) ARPES have been recently widely used in revealing the novel spin and orbital texture of the topological surface states
of three-dimensional topological insulators
~\cite{TI_REV_Kane_2010RMP,TI_REV_QiXL_2011RMP}.

The spin-resolved ARPES for surface states of \bise family of topological
insulator materials have been well studied theoretically and experimentally in
recent years, which show that the spin orientation of photo-electrons can be
completely different from their initial states and strongly depends on the
polarization of the incident light
~\cite{McIver_NN,ARPES_Bi2Se3_Fu,Theory_ARPES_StevenL_2012PRL,2013Nat.Phys,Bi2Se3_CD,
REV_CD_ARPES_Wang_2013PSS,2014PhRvX,Zhu_PRL}.
These results reveals that three different definition of the spin texture 
for the topological surface states, namely the pseudo spin and real spin
orientation for electronic states inside the crystal and that of the 
photo-emitted electrons in the vacuum, which are often mentioned within the
context of spin resolved ARPES experiments, are indeed very different and
should be clarified rigorously and studied separately
\cite{Theory_ARPES_StevenL_2012PRL,2014PhRvX}. 
In the present paper, by comparing the above mentioned three different type of ``spin texture" for the surface states of  two well known topological 
insulators, \bise and SmB$_6$, we reveal that how these three concepts are 
related to each other. 

The rest of the paper is organized as follows.
In section \ref{sec:Pseudo_spin_real_spin}, we give the formulas for the 
pseudo spin, the real spin of electrons in materials and the spin polarization 
of photo-electrons in the ARPES measurement.
Then we discuss the pseudo spin texture, the real spin texture and the 
spin-resolved and CD spectra ARPES for \bise (111) surface states in 
section \ref{sec:SARPES-bise} and for \smb (001) surface states in section
\ref{sec:SARPES-smb}. 
Conclusions are given in the end of this paper.

\section{Three different definition of ``spin texture"}
\label{sec:Pseudo_spin_real_spin}

In general, the topological surface states can be described by a 
2$\times$2 Dirac Hamiltonian 
\begin{equation}
H(\bm{k})=d_{0}(\bm{k})\sigma_{0}+d_{x}(\bm{k})\sigma_{x}+d_{y}(\bm{k})\sigma_{y}+d_{z}(\bm{k})\sigma_{z},\label{eq:h22}
\end{equation}
where $\sigma_{0}$ is identity matrix and $\sigma_{x,y,z}$ are Pauli matrices
indicating the space expanded by the eigenfunction $\varphi_{\pm}$ at the 
Dirac point.
$\varphi_{\pm}$ form Kramers doublet at the time-reversal symmetry point 
$\bm{k}=0$ and we denote them as pseudo spin in the following text.
The pseudo spin texture can be obtained by calculating the expected
value of $\bm{\sigma}$ matrix as
\begin{equation}
\langle\bm{\sigma}\rangle_{\bm{k}}=(\langle\bm{k}|\sigma_{x}|\bm{k}\rangle,\langle\bm{k}|\sigma_{y}|\bm{k}\rangle,\langle\bm{k}|\sigma_{z}|\bm{k}\rangle),\label{eq:pseudo_spin_vec}
\end{equation}
where $|\bm{k}\rangle$ is the eigenstates of Eq. (\ref{eq:h22}).

In order to get the real spin vector on the surface states, one need to know 
the real spin operator $\bm{s}$ for the surface states. The connection between 
real spin operator $\bm{s}$ and pseudo spin operators $\bm{\sigma}$ are 
characterized by the following ``g-factor'' matrix
\begin{equation}
(s_{x},s_{y},s_{z})=(\sigma_{x},\sigma_{y},\sigma_{z})\left[\begin{array}{ccc}
g_{xx} & g_{xy} & g_{xz}\\
g_{yx} & g_{yy} & g_{yz}\\
g_{zx} & g_{zy} & g_{zz}
\end{array}\right].
\label{eq:g-factor}
\end{equation}
This ``g-factor'' matrix can be obtained by projecting the real
spin operators into the surface states subspace $\varphi_{\pm}$~\cite{LiuCX_PRB_BiSe_model}. 
After obtaining $\bm{s}$, one can get the 
expectation value of real spin for any electronic state with momentum $\bm k$ as
\begin{equation}
\langle\bm{s}\rangle_{\bm{k}}=(\langle\bm{k}|s_{x}|\bm{k}\rangle,\langle\bm{k}|s_{y}|\bm{k}\rangle,\langle\bm{k}|s_{z}|\bm{k}\rangle).\label{eq:real_spin_vec}
\end{equation}

Before discussing the spin polarization of photo-electrons, we first give
some formulas for calculating the photo-emission final states. 
We start from a microscopic Hamiltonian for a system with spin-orbit
coupling, which reads
\begin{equation}
H=\frac{\bm{p}^{2}}{2m}+V(\bm{r})+\frac{\hbar}{4m^{2}c^{2}}\bm{p}\times\nabla V\cdot\bm{s},\label{eq:H_p}
\end{equation}
where $\bm{p}$ is momentum operator, $V(\bm{r})$ is crystal
potential and $\bm{s}$ is electron spin operator. The Hamiltonian
for system coupling to an electromagnetic field is obtained via the
Peierls substitution $\bm{p}\rightarrow\bm{p}-e\bm{A}$,
where $\bm{A}$ is the vector potential for the incident light. 
The linear and circular
polarized incident light are schematically shown in Fig.~\ref{fig:setup}
and their formulas are given in appendix A.
The electron-photon interaction term can then be obtained as
\begin{align}
H_{int} & =H(\bm{p}-e\bm{A})-H=-\bm{\bm{A \cdot  \mathcal{P}}}\nonumber \\
 & =-\bigg[\frac{1}{2}(A_{-}\mathcal{P}_{+}+A_{+}\mathcal{P}_{-})+A_{z}\mathcal{P}_{z}\bigg],\label{eq:H_int}
\end{align}
where $\bm{\mathcal{P}}=\frac{e}{m}\bm{p}-\frac{\hbar}{4m^{2}c^{2}}\nabla V\times\bm{s}$,
$\mathcal{P}_{\pm}=\mathcal{P}_{x}\pm i\mathcal{P}_{y}$ and $A_{\pm}=A_{x}\pm iA_{y}$.

\begin{figure}
\includegraphics[width=0.985\columnwidth]{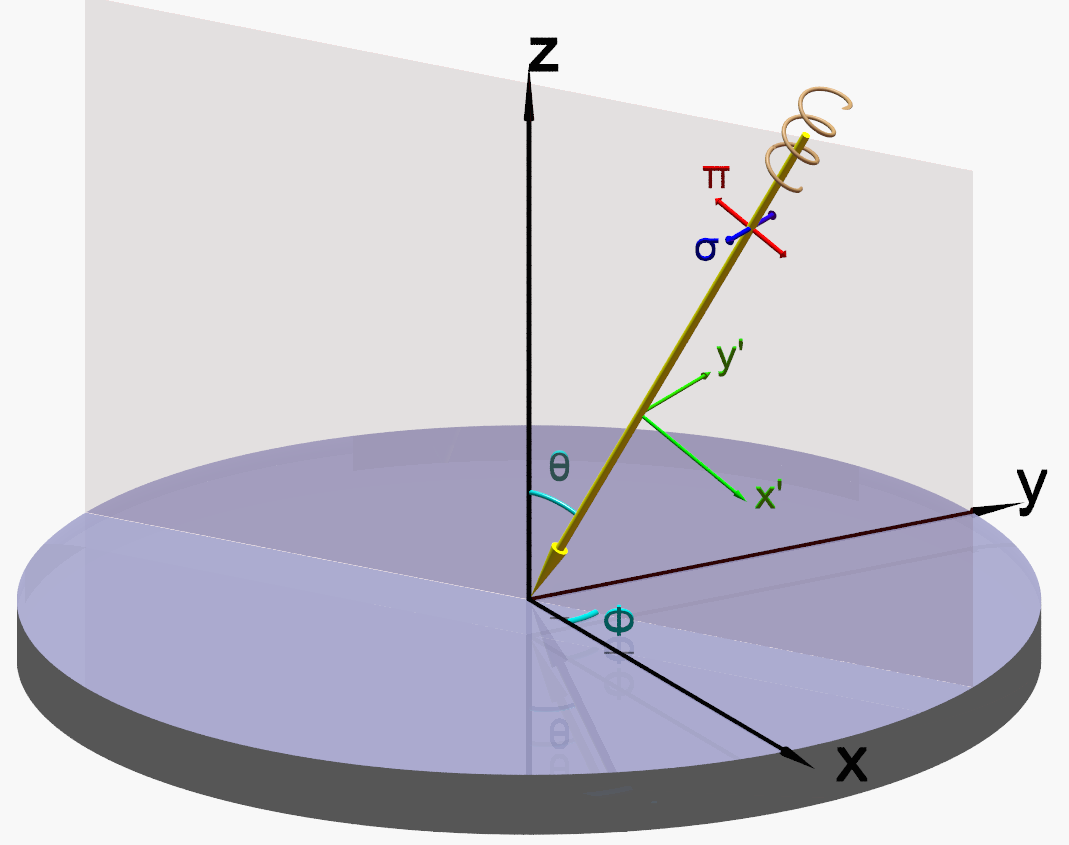}
\caption{\label{fig:setup}Diagram of the experimental geometry. Linear polarization
($\sigma$-polarized and $\pi$-polarized) and circular polarization
(right and left circular polarized) of photons can be continuously 
rotated by $\theta$ and $\phi$ angle as shown in the figure. }
\end{figure}

The photo-emitted final states can be expressed as
\begin{equation}
|f\rangle = \sum_{\alpha=\uparrow,\downarrow}|f_{\alpha}\rangle\langle f_{\alpha}|\bm{\mathcal{A}\cdot\mathcal{P}}|\bm{k}\rangle\label{eq:final_states}
\end{equation}
where $|f_{\alpha}\rangle$ are basis functions for final states with spin
$\alpha=\uparrow,\downarrow$ and $\bm{\mathcal{A}}$
is the Fourier transform of $\bm{A}$. 
The spin polarization for $|f\rangle$ states can be calculated as
\begin{equation}
\langle\bm{\tau}\rangle_{f}=(\langle f|\tau_{x}|f\rangle,\langle f|\tau_{y}|f\rangle,\langle f|\tau_{z}|f\rangle),\label{eq:tau}
\end{equation}
where $\tau_{x,y,z}$ are Pauli matrices defined in $|f_{\uparrow}\rangle$
and $|f_{\downarrow}\rangle$ space. As expressed in Eq.~(\ref{eq:final_states})
and Eq.~(\ref{eq:tau}), the spin polarization of photo-electrons
are related to the matrix elements $\langle f_{\alpha}|\bm{\mathcal{A}\cdot\mathcal{P}}|\bm{k}\rangle$,
which can be determined by considering the symmetry properties of the vector potential $\mathcal{A}$ and the
initial states $|\bm{k}\rangle$. 
In the following two section, we will discuss theoretically the 
spin polarization of the photo-emitted final states
for \bise and \smb surface states with different types of polarized 
incident light, which can be measured from spin-resolved and CD ARPES experiment.

\section{Spin-resolved and CD ARPES for \bise surface states
\label{sec:SARPES-bise}}

\begin{figure}[h]
\begin{centering}
\includegraphics[width=0.4\columnwidth]{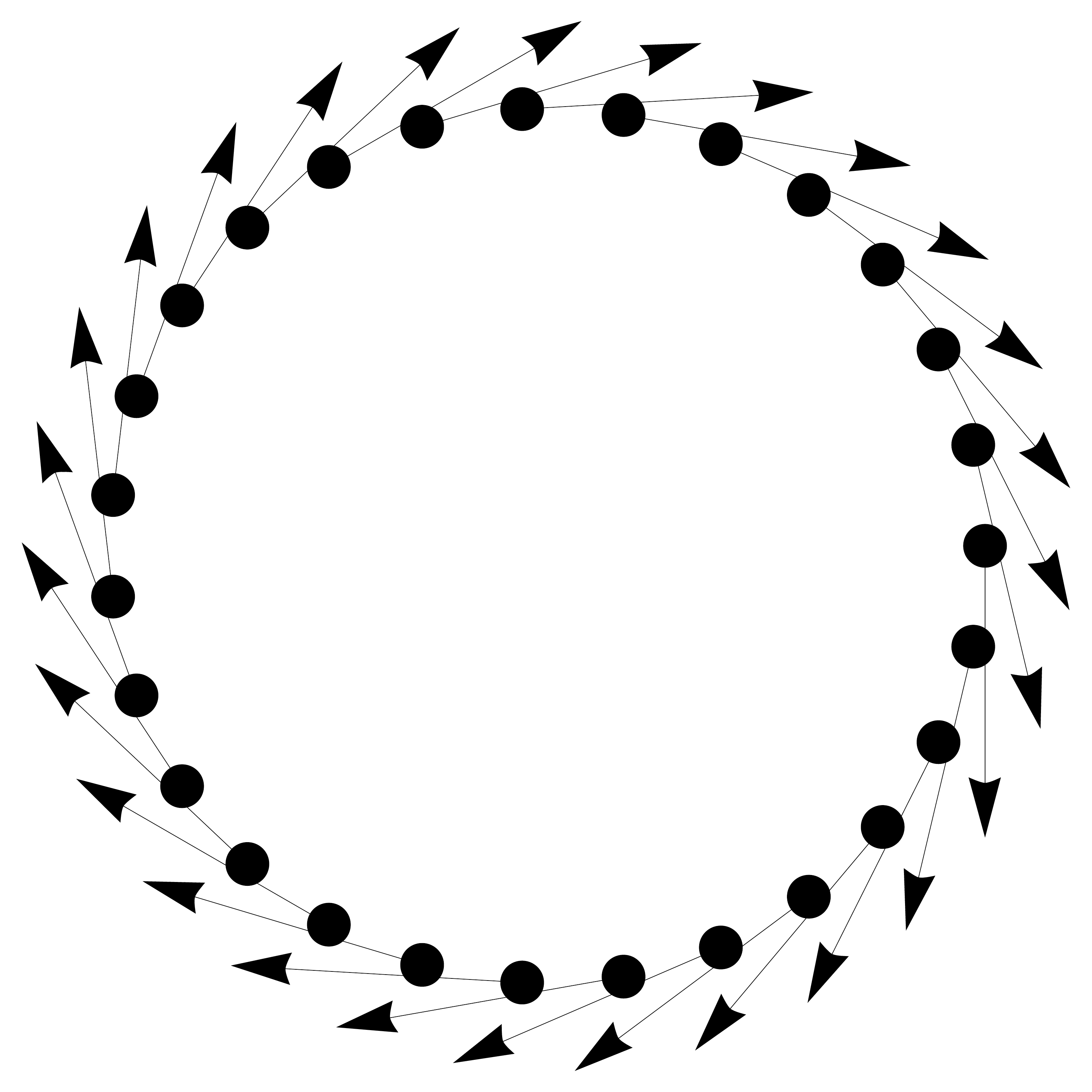}
\par\end{centering}

\protect\caption{\label{fig:bise_SST}Real spin texture of \bise (111) surface states with
energy above the Dirac cone.}
\end{figure}

For \bise family of materials with surface terminated in (111) direction,
a Dirac-like surface states exist at the $\bar{\Gamma}$ point in the surface
Brillouin zone~\cite{BiSe_ZhangHJ_2009NP,BiSe_TI_exp_Hasan_2009,BiSe_TI_exp_SZX_2010,BiSe_exp_Yazdani_2_2011}. 
A 2$\times$2 k$\cdot$p model Hamiltonian, in basis of $\varphi_{\pm}$, preserving time-reversal and 
$C_{3v}$ crystalline symmetry has been derived to describe these surface states. 
By considering the $C_{3}$ rotation symmetry, the basis functions $\varphi_{\pm}$ 
can be classified into two classes,
namely $j_{z}=\pm 1/2$ and $j_{z}=\pm 3/2$ classes. 
Other states with higher $j_{z}$ values can be reduced into 
$j_{z}=\pm 1/2$ or $j_{z}=\pm 3/2$ classes by modulating 3 because of the 
discreet $C_{3}$ rotation symmetry in crystal.
The first-principles calculations for \bise show that the pseudo spin states $\varphi_{\pm}$ transform as vectors with angular momentum $j_{z}=\pm 1/2$ and the model Hamiltonian for the \bise (111) surface states takes the following Dirac-type formula
\begin{eqnarray}
H & = & \hbar v(k_{y}\sigma_{x}-k_{x}\sigma_{y})=\hbar vk(sin\beta\sigma_{x}-cos\beta\sigma_{y})\label{eq:H_bise_kp}
\end{eqnarray}
in basis of \{$|j_{z}=+\frac{1}{2}\rangle$, $|j_{z}=-\frac{1}{2}\rangle$\}, where
$\beta$ is the angle between $\bm{k}$ and the $+\vec{k}_{x}$
direction and \textbf{$\sigma_{x,y}$} are the Pauli matrices in the pseudo
spin space. The eigenvalues for above Hamiltonian are given as $E_{n,k}=n\hbar vk$,
where $n=\pm1$ indicate the eigenvalues above and below the Dirac
cone. The Bloch periodic eigenstates are given as
\begin{equation}
|\bm{k}\rangle_{n}=u_{nk}|+\frac{1}{2}\rangle+v_{nk}|-\frac{1}{2}\rangle=\frac{1}{\sqrt{2}}\left[\begin{array}{c}
nie^{-i\beta}\\
1
\end{array}\right],\label{eq:nk_bise_kp}
\end{equation}
where $u_{nk}=nie^{-i\beta}/\sqrt{2}$ and $v_{nk}=1/{\sqrt{2}}$.
Using Eq.~(\ref{eq:pseudo_spin_vec}), the pseudo spin vectors can be calculated as
\begin{equation}
\langle\bm{\bm{\sigma}}\rangle_{\bm{k},n}\propto n(sin\beta,\;-cos\beta,\;0).
\label{eq:psuedo_spintexture_bise}
\end{equation}
The spin operator $\bm{s}$ is related to the pseudo spin operator
$\bm\sigma$ as $(s_{x},s_{y},s_{z})=(g_{xx}\sigma_{x},g_{yy}\sigma_{y},g_{zz}\sigma_{z})$, whit $g_{xx,yy,zz}$ to be some constants \cite{LiuCX_PRB_BiSe_model}. The real spin is
proportional to the pseudo spin which can be calculated as
\begin{equation}
\langle\bm{{s}}\rangle_{\bm{k},n}\propto 
n(g_{xx}sin\beta,\;-g_{yy}cos\beta,\;0)\label{eq:spintexture_bise}
\end{equation}
as shown in Fig. \ref{fig:bise_SST}.

Substitute Eq.~(\ref{eq:nk_bise_kp}) into Eq.~(\ref{eq:final_states}),
the photo-emitted final states can be expressed as
\begin{eqnarray}
|f\rangle & \propto & |+\frac{1}{2}\rangle_{ff}\langle+\frac{1}{2}|\bm{\mathcal{A}\cdot\mathcal{P}}|+\frac{1}{2}\rangle_{i}u_{nk}\nonumber \\
 & + & |+\frac{1}{2}\rangle_{ff}\langle+\frac{1}{2}|\bm{\mathcal{A}\cdot\mathcal{P}}|-\frac{1}{2}\rangle_{i}v_{nk}\nonumber \\
 & + & |-\frac{1}{2}\rangle_{ff}\langle-\frac{1}{2}|\bm{\mathcal{A}\cdot\mathcal{P}}|+\frac{1}{2}\rangle_{i}u_{nk}\nonumber \\
 & + & |-\frac{1}{2}\rangle_{ff}\langle-\frac{1}{2}|\bm{\mathcal{A}\cdot\mathcal{P}}|-\frac{1}{2}\rangle_{i}v_{nk}.\label{eq:final_states_bise}
\end{eqnarray}
where the subscript $i$ and $f$ indicate the initial and final states respectively.
The matrix elements in Eq.~(\ref{eq:final_states_bise}) can be determined
by the symmetry considerations as discussed below.

At $\bar{\Gamma}$ point the symmetry is characterized by $C_{3v}$
crystalline symmetry, which is reduced from space group $R3\bar{m}$
in the present of (111) direction surface and consists of a threefold
rotation $C_{3}$ around $z$ axis and a mirror operation $M_{x}$:
$x\rightarrow-x$. Under these two operations, the basis functions
are transformed as follows
\begin{eqnarray}
M_{x}|\pm\frac{1}{2}\rangle_{i,f} & = & i\;|\mp\frac{1}{2}\rangle_{i,f},\label{eq:Mx_bise}\\
C_{3}|\pm\frac{1}{2}\rangle_{i,f} & = & e^{-i\frac{2\pi}{3}\times(\pm\frac{1}{2})}|\pm\frac{1}{2}\rangle_{i,f}.\label{eq:C3_bise}
\end{eqnarray}
With the properties shown in Eq.~(\ref{eq:Mx_bise}) and Eq.~(\ref{eq:C3_bise}),
only the following four matrix elements in Eq.~(\ref{eq:final_states_bise})
are nonzero
\begin{eqnarray}
_{f}\langle+\frac{1}{2}|\mathcal{P}_{+}|-\frac{1}{2}\rangle_{i}= & -{}_{f}\langle-\frac{1}{2}|\mathcal{P}_{-}|+\frac{1}{2}\rangle_{i} & =a,\label{eq:-6}\\
_{f}\langle+\frac{1}{2}|\mathcal{P}_{z}|+\frac{1}{2}\rangle_{i}= & _{f}\langle-\frac{1}{2}|\mathcal{P}_{z}|-\frac{1}{2}\rangle_{i} & =c,\label{eq:bise_pz}
\end{eqnarray}
where we used the property of $C_{3z}\mathcal{P}_{\pm}C_{3z}^{\dagger}=e^{\mp i\frac{2\pi}{3}}\mathcal{P}_{\pm}$
and $a$, $c$ are complex parameters that should be determined from
the first-principles calculations or by fitting with experimental
results. 
With the help of Eq. (\ref{eq:-6}) and Eq. (\ref{eq:bise_pz}), the finial states $|f\rangle$ can
thus be rewritten as
\begin{eqnarray}
|f\rangle & \propto & \big(av_{nk}\mathcal{A}_{-}+cu_{nk}\mathcal{A}_{z}\big)|+\frac{1}{2}\rangle_{f}\nonumber \\
 & + & \big(cv_{nk}\mathcal{A}_{z}-au_{nk}\mathcal{A}_{+}\big)|-\frac{1}{2}\rangle_{f}\nonumber \\
 & = & \frac{1}{\sqrt{2}}\left[\begin{array}{c}
a\mathcal{A}_{-}+icne^{-i\beta}\mathcal{A}_{z}\\
-iane^{-i\beta}\mathcal{A}_{+}+c\mathcal{A}_{z}
\end{array}\right].\label{eq:final_states_bise_A}
\end{eqnarray}

The spin polarization of final states $|f\rangle$ can be calculated for light with different types of polarization.

\begin{figure}[H]
\centering{}\includegraphics[width=0.99\columnwidth]{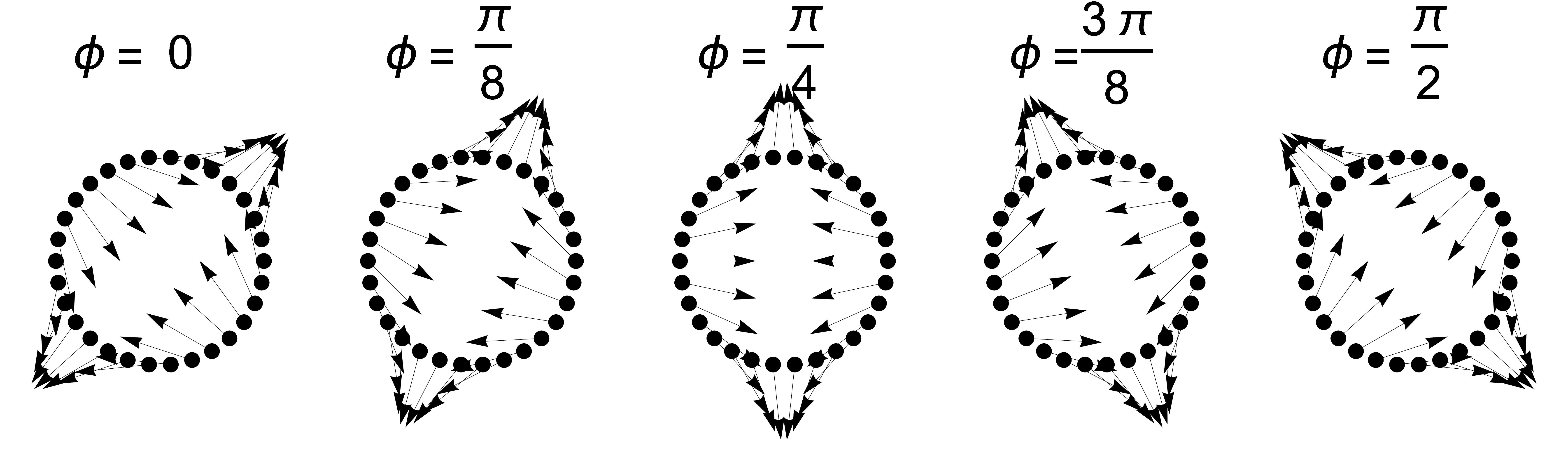}
\caption{Spin polarization of photon-electrons with
$\sigma$ polarized light. Arrows indicate the spin directions in
the $xy$ plane.}\label{fig:bise_s.pol}
\end{figure}

(i) For $\sigma$-polarized light, $\bm{\mathcal{A}}_{\sigma}=A_{0}(-sin\phi,\;cos\phi,\;\mbox{0})$, Eq.~(\ref{eq:final_states_bise_A}) takes the formula as
\begin{eqnarray}
|f\rangle & \propto & \frac{aA_{0}}{\sqrt{2}}\left[\begin{array}{c}
-sin\phi-icos\phi\\
-ine^{-i\beta}(-sin\phi+icos\phi)
\end{array}\right]\nonumber \\
 & \propto & \frac{aA_{0}}{\sqrt{2}}\left[\begin{array}{c}
1\\
ine^{-i(\beta-2\phi)}
\end{array}\right].
\end{eqnarray}
The spin
polarization of final states $|f\rangle$ are calculated as
\begin{equation}
\langle\bm{\tau}\rangle_{f}\propto n\big[sin(\beta-2\phi),\;cos(\beta-2\phi),\;0\big].\label{eq:-20}
\end{equation}
The spin polarization with different value of azimuth angle $\phi$ are shown in Fig.~\ref{fig:bise_s.pol} and we set the energy of the initial states above the Dirac cone in the following text.

\begin{figure}[H]
\includegraphics[clip,width=0.95\columnwidth]{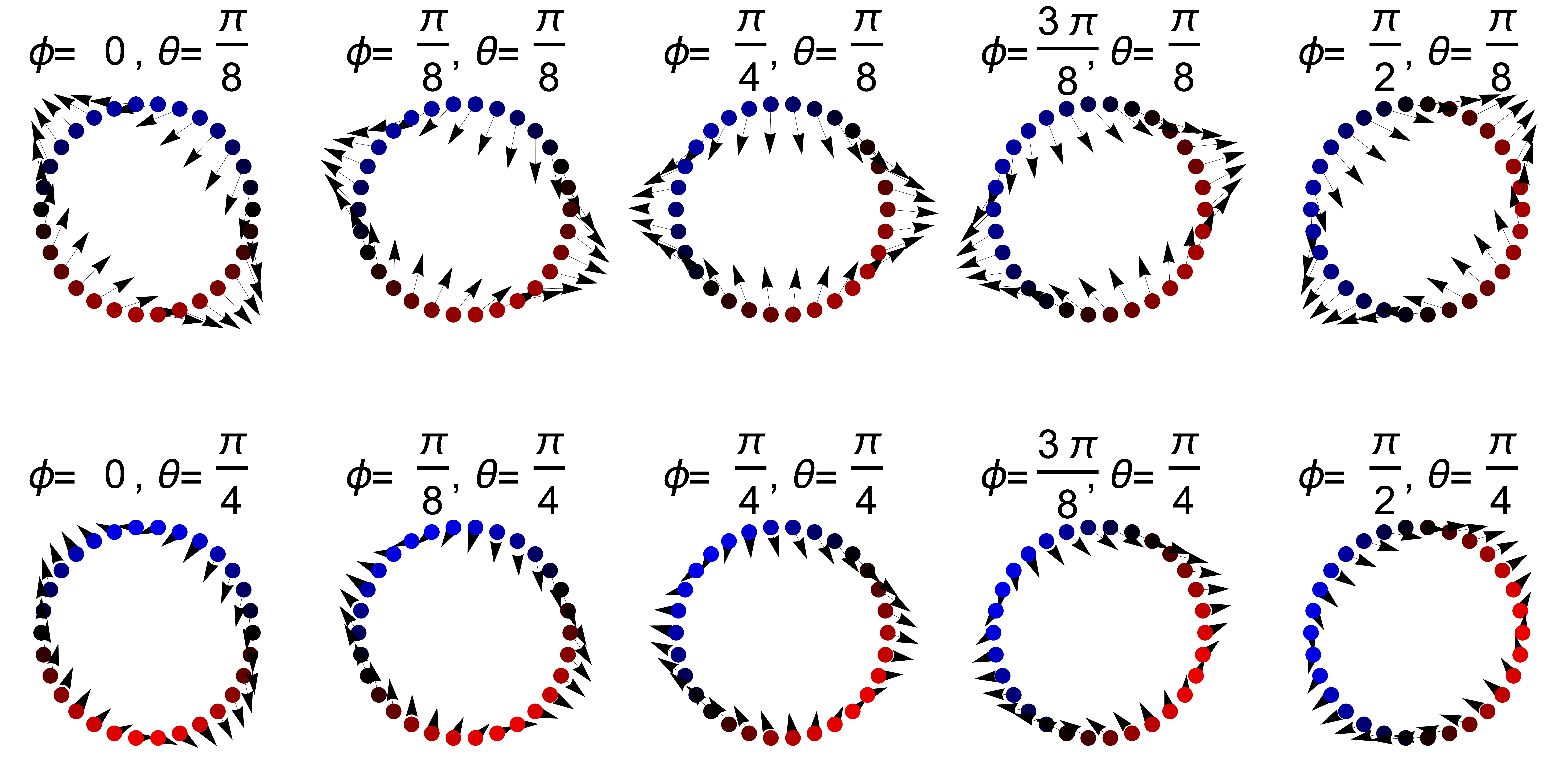} \protect\caption{\label{fig:bise_p.pol}(Color online) Spin polarization of photon-electrons with $\pi$-polarized light. Arrows indicate the spin component in the $xy$ plane. 
Color indicate the spin $z$ component: 
red for $\langle \tau_{z}\rangle<0$,
blue for $\langle \tau_{z}\rangle>0$ and 
black for $\langle \tau_{z}\rangle=0$.}
\end{figure}

(ii) For $\pi$-polarized light, we have 
$\bm{\mathcal{A}_{\pi}}=A_{0}(cos\theta cos\phi,\;cos\theta sin\phi,\;-sin\theta)$.
Substitute $\mathcal{A}_{\pi}$ into Eq.~(\ref{eq:final_states_bise_A}), we obtain the spin polarization of the photo-electrons as
\begin{eqnarray}
\langle\bm{\tau}_{x}\rangle_{f} & = & -2n|a|^{2}cos^{2}\theta sin(\beta-2\phi)+2n|c|^{2}sin\beta sin^{2}\theta\nonumber \\
 &  & -i(a^{*}c-ac^{*})sin2\theta sin\phi,\\
\langle\bm{\tau}_{y}\rangle_{f} & = & -2n|a|^{2}cos^{2}\theta cos(\beta-2\phi)-2n|c|^{2}cos\beta sin^{2}\theta\nonumber \\
 &  & +i(a^{*}c-ac^{*})sin2\theta cos\phi,\\
\langle\bm{\tau}_{z}\rangle_{f} & = & -n(a^{*}c+ac^{*})sin2\theta sin(\beta-\phi),\label{eq:-40}
\end{eqnarray}
and the spin polarization textures with parameters 
$a=-0.9 + 0.1i$, $c=0.5 - 0.1i$ with unit $\rm A\cdot m$
are shown in Fig.~\ref{fig:bise_p.pol}.

(iii) For circular polarized light,
\textbf{ $\bm{\mathcal{A}}_{\eta}=A_{0}(cos\theta cos\phi+\eta isin\phi,\;cos\theta sin\phi-\eta icos\phi,\;-sin\theta)$},
where $\eta=\pm1$ are the index for right/left-handed circular polarized light. Substitute $\mathcal{A_{\eta}}$
into Eq.~(\ref{eq:final_states_bise}), the spin polarization for photo-electrons
are calculated as
\begin{eqnarray}
\langle\bm{\tau}_{x}\rangle_{f} & = & sin\theta\bigg(-\eta(a^{*}c+ac^{*})cos\phi+i(a^{*}c-ac^{*})cos\theta sin\phi\nonumber \\
 &  & +n|c|^{2}sin\beta sin\theta-n|a|^{2}sin\theta sin(2\phi-\beta)\bigg),\\
\langle\bm{\tau}_{y}\rangle_{f} & = & sin\theta\bigg(-\eta(a^{*}c+ac^{*})sin\phi-i(a^{*}c-ac^{*})cos\theta cos\phi\nonumber \\
 &  & -n|c|^{2}cos\beta sin\theta+n|a|^{2}sin\theta cos(2\phi-\beta)\bigg),\\
\langle\bm{\tau}_{z}\rangle_{f} & = & -2\eta|a|^{2}cos\theta-\eta ni(a^{*}c-ac^{*})sin\theta cos(\beta-\phi)\nonumber \\
 &  & +n(a^{*}c+ac^{*})sin\theta cos\theta sin(\beta-\phi).
\end{eqnarray}

\begin{figure}[H]
\begin{centering}
\includegraphics[width=0.45\columnwidth]{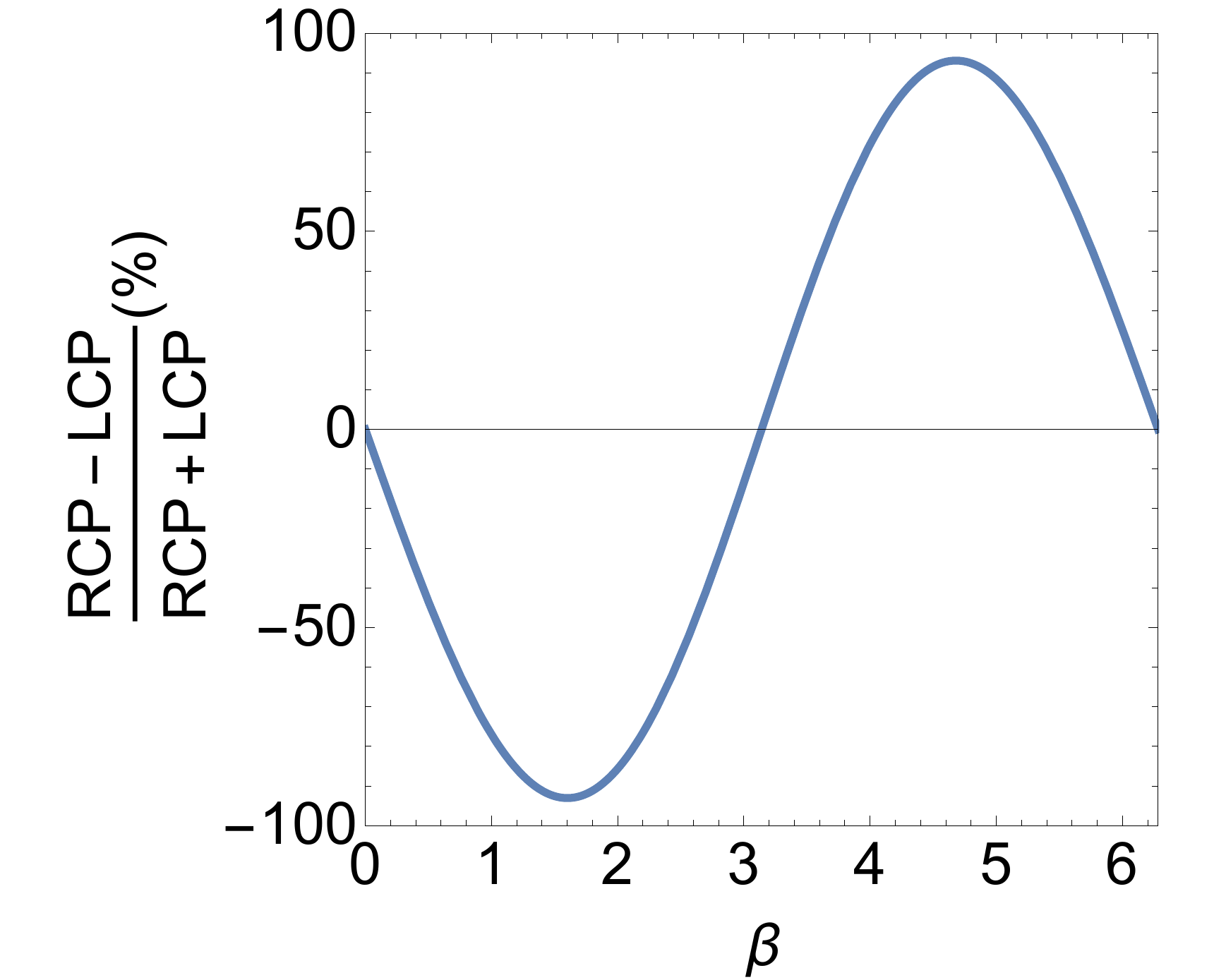} \includegraphics[width=0.45\columnwidth]{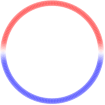}
\par\end{centering}

\protect\caption{\label{fig:bise_CD}(Color online) The calculated CD-ARPES spectra for \bise (111) surface states with $\theta=3\pi/8$ and $\phi=0$.}
\end{figure}

Experimentally, the CD-ARPES is an alternative method for probing
the spin texture of topological surface states. The CD value
is defined by taking the difference of photo-emission transition rate
for photon with opposite helicity. The photo-emission transition rate
are expressed as
\begin{align}
I_{\eta} & \propto \sum_{\sigma}|\langle f_{\sigma}|\mathcal{A}_{\eta}\cdot\mathcal{P}|k\rangle|^{2}\nonumber \\
 & =\frac{1}{4}(|a|^{2}+|a|^{2}cos^{2}\theta+4|c|^{2}sin^{2}\theta\nonumber \\
 & +2nIm[a^{*}c]cos(\beta-\phi)sin2\theta\nonumber \\
 & +2n\eta Re[a^{*}c]sin\theta sin(\beta-\phi)),
 \label{eq:I_eta}
\end{align}
where $Im$ and $Re$ refers to the imaginary and real operators. 
The CD-ARPES spectra can thus be calculated as
\begin{align}
I_{CD} & =\frac{I_{R}-I_{L}}{I_{R}+I_{L}}\nonumber \\
 & =4nRe[a^{*}c]sin\theta sin(\beta-\phi)/\bigg(|a|^{2}(1+cos^{2}\theta)\nonumber \\
 & +4|c|^{2}sin^{2}\theta+2nIm[a^{*}c]sin2\theta cos(\beta-\phi)\bigg)\label{eq:I_CD_bise}.
\end{align}
The calculated $I_{CD}$ for \bise (111) surface states are shown in
Fig.~\ref{fig:bise_CD}.

\section{Spin-resolved and CD-ARPES for SmB$_6$ surface
states\label{sec:SARPES-smb}}

Recently the mixed valence compound \smb has been proposed to be
a topological insulator and attracts lots of research interest\cite{SmB6_KTI_Dzero_2010PRL,SmB6_KTI_Dzero_2010PRL,SmB6_Dzero_prb_2012,SmB6_lu_smb6_2013,SmB6_Dzero_prl_2013,SmB6_Dzero_JETP_2013}.
Unlike the \bise family of materials, the strong correlation effects
in mixed valence topological insulators are crucial in understanding electronic structure
owing to the partially filled 4f bands. There are two main effects
induced by the on-site Coulomb interaction among the 4f-electrons:
the strong modification of the 4f band width and the correction to the 
effective spin orbit coupling and crystal field.
As a consequence, the band inversion in the modified band structure
happens between 5d and 4f band around three X points at the BZ boundary.
If a surface terminated in the (001) direction, one X point projects
to the $\bar{\Gamma}$ points on the surface BZ and the other two
X points project to $\bar{Y}$ and $\bar{X}$ points and leading to
three different Dirac points on the (001) surfaces. 
In this section, we will discuss the relations between the above mentioned 
three spin textures and calculate the CD spectrum for the 
surface states near $\bar{\Gamma}$, $\bar{Y}$ and $\bar{X}$ points.

\subsection{Surface states at $\bar{\Gamma}$ point}

The crystalline symmetry at $\bar{\Gamma}$ point is characterized
by double group of $C_{4v}$. From group theory we know that there
are two kinds of two-dimensional irreducible representations for the
double group of $C_{4v}$, which are $j_{z}=\pm 1/2$ and
$j_{z}=\pm 3/2$ representation respectively. 
The first-principles calculations show that the Dirac surface states 
at $\bar{\Gamma}$ point belong to the representation of $j_{z}=\pm3/2$. 
Therefore the effective Hamiltonian for surface states 
near $\bar{\Gamma}$ point is given 
as \cite{SmB6_modelH_Roy_2014PRB,YuRui_SmB_NJP}
\begin{equation}
H_{\bar{\Gamma}}=-\hbar v(k_{y}\sigma_{x}+k_{x}\sigma_{y})=-\hbar vk(sin\beta\sigma_{x}+cos\beta\sigma_{y})\label{eq:-21-1}
\end{equation}
in the basis of $\{|j_{z}=+\frac{3}{2}\rangle,|j_{z}=-\frac{3}{2}\rangle\}$, 
where $\beta$ is the angle between k and the $+k_{x}$ direction
and $\sigma_{x,y}$ are the Pauli matrices in the pseudo spin space 
expand by $|j_{z}=\pm\frac{3}{2}\rangle$.

\begin{figure}[H]
\begin{centering}
\includegraphics[width=0.7\columnwidth]{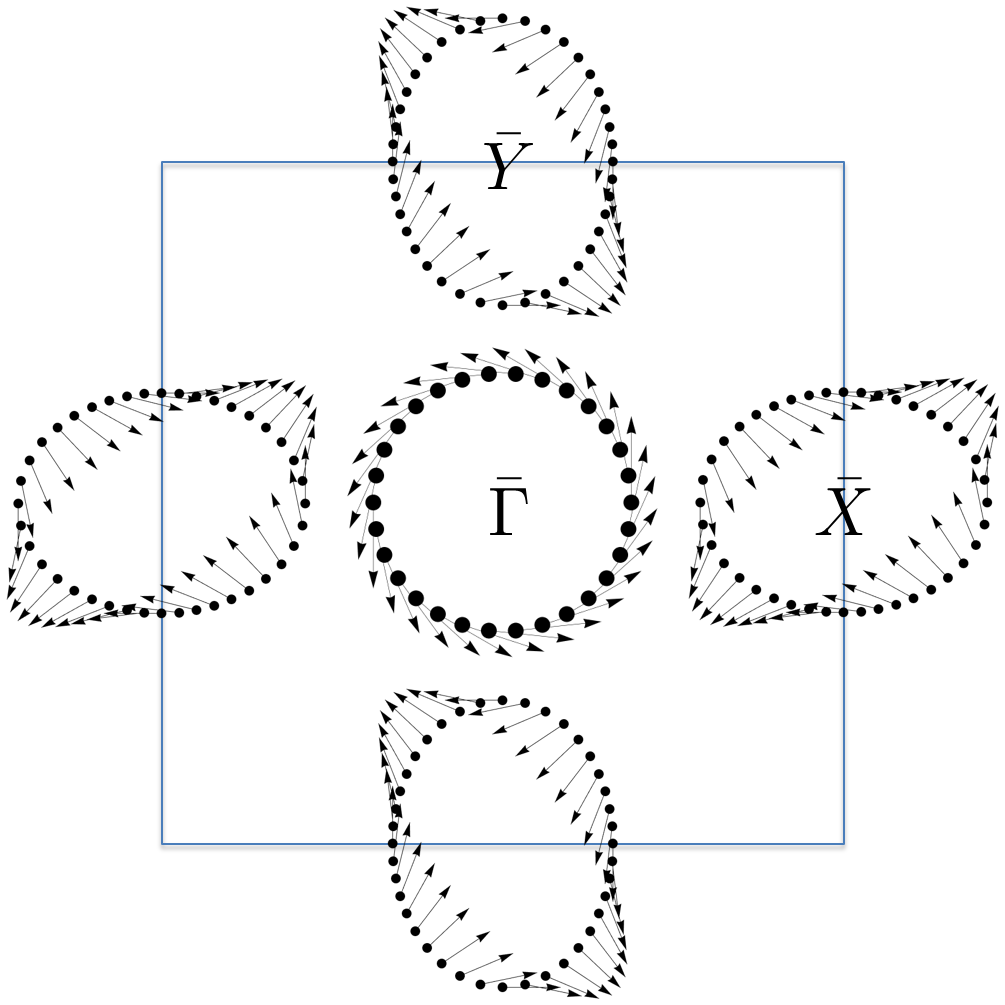}
\par\end{centering}

\protect\caption{\label{fig:SmB6_SST} Real spin texture of the \smb (001) surface states
near $\bar{\Gamma}$, $\bar{Y}$ and $\bar{X}$ point with energy
above the Dirac cone. }
\end{figure}

The eigenvalue and Bloch periodic eigenstate near $\bar{\Gamma}$
point are $E_{n}=n\hbar vk$ and
\begin{equation}
|\bm{k}\rangle_{n}=\frac{1}{\sqrt{2}}\bigg(nie^{i\beta}|+\frac{3}{2}\rangle_{i}+|-\frac{3}{2}\rangle_{i}\bigg)=\frac{1}{\sqrt{2}}\left[\begin{array}{c}
ine^{i\beta}\\
1
\end{array}\right].\label{eq:smb6_GM_eigenstates}
\end{equation}
The pseudo spin texture are calculated as
\begin{equation}
\langle\bm{\sigma}\rangle_{\bm{k},n}\propto -n(sin\beta,\;cos\beta,\;0).
\end{equation}
The ``g-faoctor'' connecting real spin operator $\bm{s}$ and
pseudo spin operator $\bm{\sigma}$ in Eq.~(\ref{eq:g-factor})
are given as~\cite{YuRui_SmB_NJP} $g_{xx}=0.095$, $g_{yy}=-0.095$,
$g_{zz}=0.068$. Then the real spin vectors for the TSS are given by
\begin{equation}
\langle{\bm{s}}\rangle_{\bm{k},n}\propto -n(g_{xx}sin\beta,\;g_{yy}cos\beta,\;0)\label{eq:-5-1-1}
\end{equation}
as shown in Fig.~\ref{fig:SmB6_SST}.

The $C_{4v}$ symmetry contains $C_{4}$
rotation symmetry and mirror symmetry $M_{x}$. The basis function
$|\pm\frac{3}{2}\rangle_{i}$ for initial states and $|\pm\frac{1}{2}\rangle_{f}$
for final states satisfy the following symmetry properties
\begin{eqnarray}
M_{x}|\pm\frac{3}{2}\rangle_{i} & = & -i|\mp\frac{3}{2}\rangle_{i},\label{eq:-25}\\
C_{4}|\pm\frac{3}{2}\rangle_{i} & = & e^{-i\frac{2\pi}{4}\times(\pm\frac{3}{2})}|\pm\frac{3}{2}\rangle_{i},\label{eq:-2-1}\\
C_{4}|\pm\frac{1}{2}\rangle_{f} & = & e^{-i\frac{2\pi}{4}\times(\pm\frac{1}{2})}|\pm\frac{1}{2}\rangle_{f}.\label{eq:-16}
\end{eqnarray}
With the similar arguments used in the previous section, the nonzero matrix
elements in Eq.~(\ref{eq:final_states}) are obtained as listed below
\begin{equation}
_{f}\langle+\frac{1}{2}|\mathcal{P}_{-}|+\frac{3}{2}\rangle_{i}={}_{f}\langle-\frac{1}{2}|\mathcal{P}_{+}|-\frac{3}{2}\rangle_{i}=a.\label{eq:matrix_elements_smb6}
\end{equation}
The matrix element for $\mathcal{P}_{z}$ is vanish for that it cannot conserve the total
angular momentum along $z$ direction.
This results is different from the \bise case, where the surface
states with $j_{z}=\pm\frac{1}{2}$ lead to the nonzero matrix element for $\mathcal{P}_{z}$
as shown in Eq.~(\ref{eq:bise_pz}). Substitute Eq.~(\ref{eq:smb6_GM_eigenstates})
and Eq.~(\ref{eq:matrix_elements_smb6}) into Eq.~(\ref{eq:final_states}),
the final state is obtained as
\begin{eqnarray}
|f\rangle & \propto & \frac{a}{\sqrt{2}}\left[\begin{array}{c}
ine^{i\beta}\mathcal{A}_{+}\\
\mathcal{A}_{-}
\end{array}\right].\label{eq:final_states_smb6_GM}
\end{eqnarray}

\begin{figure}[h]
\centering{}\includegraphics[width=0.9\columnwidth]{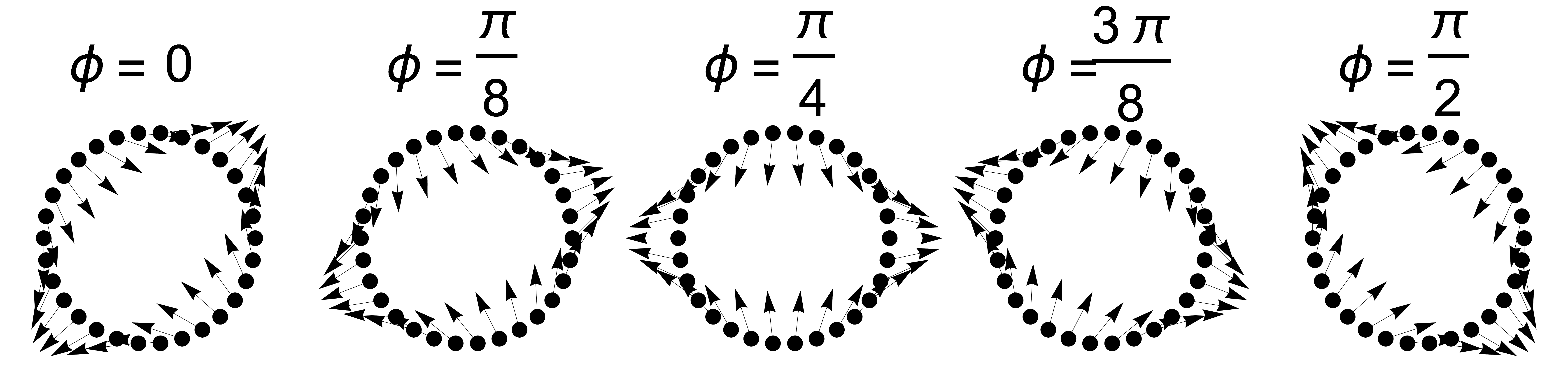}
\protect\caption{\label{fig:SmB6_G_s.pol}Spin polarization of photon-electrons with
$\sigma$ polarized light. Arrows indicate the spin component in the xy plane.}
\end{figure}

(i) For $\sigma$-polarized light, 
$\mathcal{A}_{\sigma}=A_{0}(-sin\phi,\;cos\phi,\;0)$,
the final states takes the form of
\begin{eqnarray}
|f\rangle & \propto & \frac{aA_{0}}{\sqrt{2}}\left[\begin{array}{c}
-ine^{i(\beta+2\phi)}\\
1
\end{array}\right],\label{eq:-19-1}
\end{eqnarray}
and the spin polarization for photo-electrons can be calculated as
\begin{equation}
\langle{\bm\tau\rangle_{f}\propto}n\big[sin(\beta+2\phi),cos(\beta+2\phi),0\big]\label{eq:s_sp}
\end{equation}
as shown in Fig.~\ref{fig:SmB6_G_s.pol}. The spin texture takes a
different rotation manner as tuning $\phi$ from $\phi=0$ to $\pi/2$
compare to \bise as shown in Fig.~\ref{fig:bise_s.pol}.

\begin{figure}[h]
\begin{centering}
\includegraphics[width=1\columnwidth]{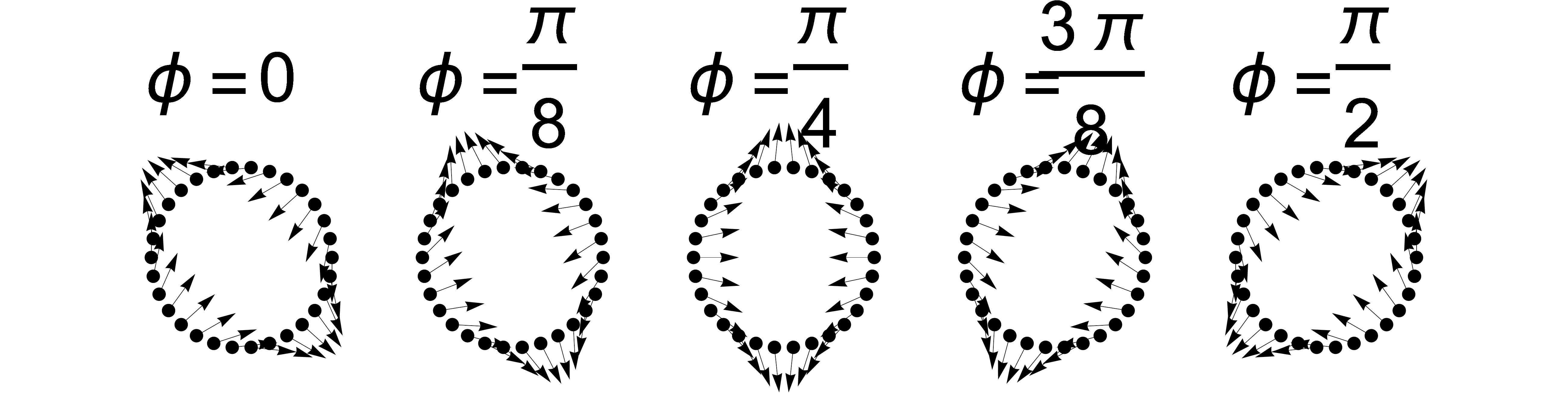}
\par\end{centering}

\protect\caption{\label{fig:SmB6_G_p.pol}Spin polarization of photon-electrons with
$\pi$-polarized light. Arrows indicate the spin component in the xy plane.}
\end{figure}

(ii) For $\pi$-polarized light, $\bm{\mathcal{A}_{\pi}}$=$A_{0}(cos\theta cos\phi,$ $\;cos\theta sin\phi,$ $\;-sin\theta)$, the final states can be calculated as
\begin{eqnarray}
|f\rangle & \propto & \propto\frac{aA_{0}cos\theta}{\sqrt{2}}\left[\begin{array}{c}
ine^{i(\beta+2\phi)}\\
1
\end{array}\right],\label{eq:-19-1-2}
\end{eqnarray}
and the spin vector for $|f\rangle$ is calculated as
\begin{equation}
\langle{\bm\tau}\rangle_{f}\propto-ncos^{2}\theta( sin(\beta+2\phi),cos(\beta+2\phi),0)\label{eq:s_ppol}
\end{equation}
as shown in Fig.~\ref{fig:SmB6_G_p.pol}. Different to the case in
\bise system, for \smb (001) surface states the $\pi$-polarized
light does not induce the $z$ direction component in the spin orientation,
for the photo-emission matrix element of $\mathcal{P}_{z}$ is vanish
under the symmetry constraint.

(iii) For circular polarized light, $\mathcal{A}_{\eta}=A_{0}(cos\theta cos\phi+i\eta sin\phi,\;cos\theta sin\phi-i\eta cos\phi,\;-sin\theta)$,
where $\eta=\pm 1$ indicate the RCP/LCP light, the final states can 
be calculated as
\begin{eqnarray}
|f\rangle & \propto & \frac{aA_{0}}{\sqrt{2}}\left[\begin{array}{c}
-in(cos\theta+\eta)e^{i(\beta+2\phi)}\\
(cos\theta-\eta)
\end{array}\right]\label{eq:}
\end{eqnarray}
with spin vector
\begin{equation}
\langle\bm{\tau\rangle}_{f}\propto(nsin^{2}\theta sin(\beta+2\phi),nsin^{2}\theta cos(\beta+2\phi),2\eta cos\theta)\label{eq:s_RLCP}
\end{equation}
The photo-emission transition rate is calculated as
\begin{equation}
I_{\eta}=\frac{|a|^{2}}{4}(1+cos^{2}\theta),
\label{eq:I_eta_smb6}
\end{equation}
which is independent with light helicity $\eta$ and lead to $I_{CD}=0$. 
The reason for obtained the vanish CD spectra is that here we only keep 
up to the zeroth order perturbation for the initial states.
Keep up to the first order perturbations terms in the initial states, 
we get~\cite{SOT_ZhangHaijun_2013PRL}
\begin{equation}
|\phi_{\pm}\rangle=a_{1}|\pm\frac{3}{2}\rangle\pm ia_{2}k_{\pm}|\pm\frac{1}{2}\rangle\pm ia_{3}k_{\mp}|\pm\frac{5}{2}\rangle,
\label{eq:basis_1k}
\end{equation}
where $a_{1,2,3}$ are material dependent parameters. The above wave
functions are constructed by considering the conservation of the total
angular momentum $j_{z}$ in $z$ direction~\cite{SOT_ZhangHaijun_2013PRL}. For example, $k_{\pm}$
carry the angular momentum $\pm1$, so the total angular momentum
in $z$ direction is $\pm1/2$ for the second and third terms in Eq.~(\ref{eq:basis_1k}).

\begin{figure}[h]
\begin{centering}
\includegraphics[width=1\columnwidth]{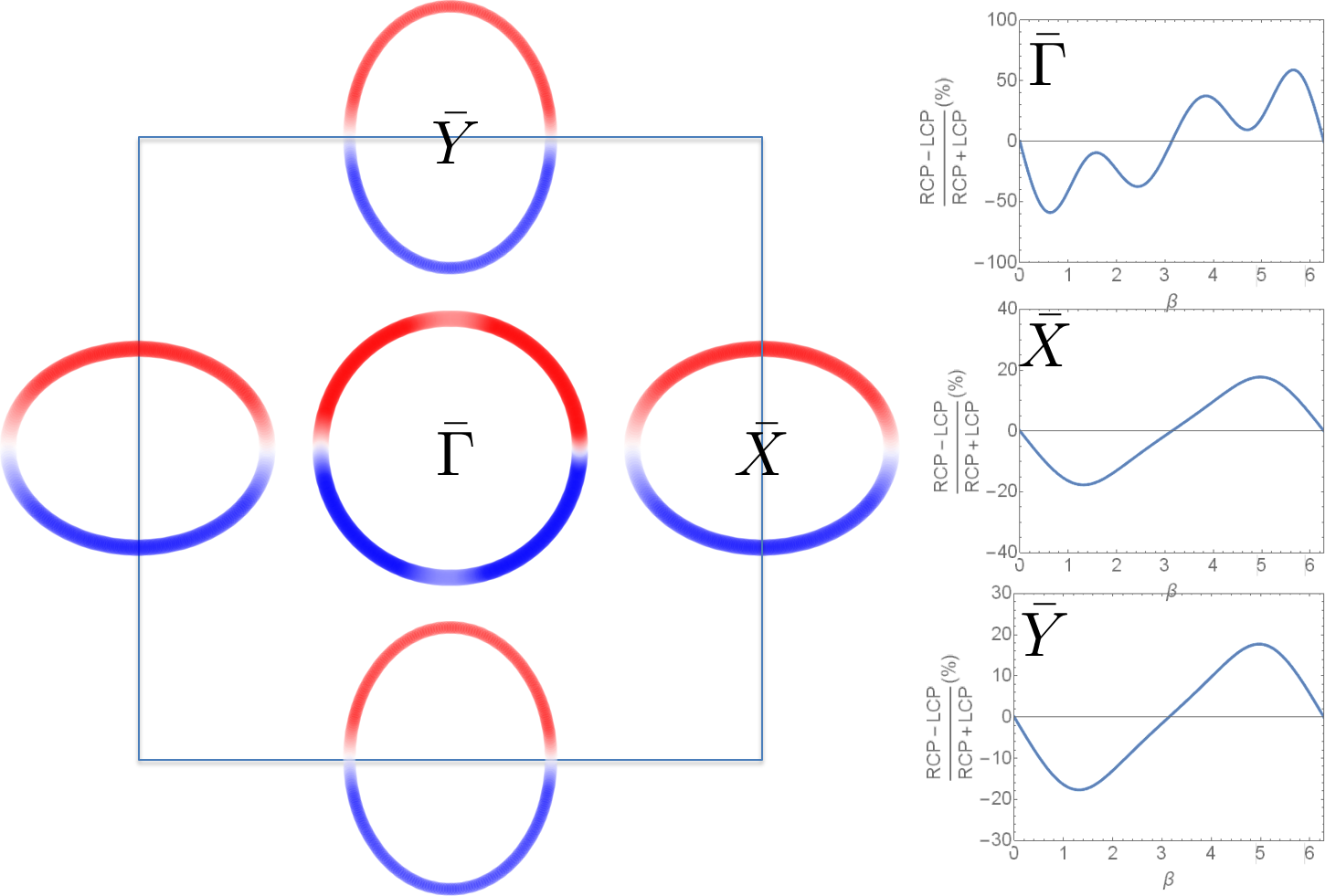}
\par\end{centering}

\protect\caption{\label{fig:SmB6_GM_CD}(Color online) Calculated CD spectra for \smb (001) surface states near $\bar{\Gamma}$, $\bar{Y}$ and $\bar{X}$ points.}
\end{figure}

Taking the symmetry consideration into the matrix elements of $\mathcal{P}$, we find
that the following terms are nonzero
\begin{equation}
_f\langle+\frac{1}{2}|\mathcal{P}_{+}|-\frac{1}{2}\rangle_i=_f\langle-\frac{1}{2}|\mathcal{P}_{-}|+\frac{1}{2}\rangle_i=c_{1},\label{eq:-3}
\end{equation}
\begin{equation}
_f\langle+\frac{1}{2}|\mathcal{P}_{-}|+\frac{3}{2}\rangle_i=_f\langle-\frac{1}{2}|\mathcal{P}_{+}|-\frac{3}{2}\rangle_i=c_{2},\label{eq:-4}
\end{equation}
\begin{equation}
_f\langle+\frac{1}{2}|\mathcal{P}_{-}|-\frac{5}{2}\rangle_i=_f\langle-\frac{1}{2}|\mathcal{P}_{+}|+\frac{5}{2}\rangle_i=c_{3},\label{eq:-5}
\end{equation}
\begin{equation}
_f\langle+\frac{1}{2}|\mathcal{P}_{z}|+\frac{1}{2}\rangle_i=_f\langle-\frac{1}{2}|\mathcal{P}_{z}|-\frac{1}{2}\rangle_i=c_{4}.\label{eq:-7}
\end{equation}
The difference of photo-emission transition rate under right- and left-handed
circular polarized light is calculated as
\begin{eqnarray}
I_{R}-I_{L}= & \bigg[Im[c_{4}c_{1}^{*}]a_{2}nkcos(3\beta+\phi)+\big(Im[c_{4}c_{2}^{*}]a_{1}\nonumber \\
 & +Im[c_{3}c_{4}^{*}]a_{3}nk\big)cos(\beta-\phi)\bigg]2a_{2}ksin\theta,
\end{eqnarray}
and the CD values around $\bar{\Gamma}$ point are shown in Fig.~\ref{fig:SmB6_GM_CD}
with parameters
$a_1=-0.25$,
$a_2=0.6$,
$a_3=-0.52$,
$c_1=-0.2-0.1i$,
$c_2=-0.6+0.1i$,
$c_3=-0.6-0.4i$,
$c_4=0.1-0.3i$ where $c_i$ with unit $\rm A\cdot m$ and these parameters qualitatively reproduce the experimental result in Ref.~\onlinecite{FengDL_SmB6_CD_ARPES_2013_NatComm}.

\subsection{Surface states at $\bar{Y}$ point}

The symmetry of Dirac type surface states locate at $\bar{Y}$ and $\bar{X}$
points are characterized by $C_{2v}$ and time reversal symmetry.
The first-principles calculations show that the eigenstates possess angular momentum $j_{z}=\pm\frac{3}{2}$ at these surface Dirac points.
The surface states Hamiltonian at $\bar{Y}$ point is given as~\cite{YuRui_SmB_NJP}
\begin{equation}
H_{\bar{Y}}=ak_{y}\sigma_{x}+bk_{x}\sigma_{y}
\label{eq:-21-1-1}
\end{equation}
in basis of \{$|j_{z}=+3/2\rangle$, $|j_{z}=-3/2\rangle$\}, where
the Pauli matrices indicate the pseudo spin space,
$\sigma_{\pm}=\sigma_{x}\pm i\sigma_{y}$,
$a$, $b$ are material dependent parameters.
The eigenvalues and Bloch periodic
eigenstates are $E_{n}=n\sqrt{a^{2}k_{y}^{2}+b^{2}k_{x}^{2}}$ and
\begin{eqnarray}
|k\rangle_{n} & = & \frac{1}{\sqrt{2}\sqrt{a^{2}k_{y}^{2}+b^{2}k_{x}^{2}}}\left[\begin{array}{c}
n\sqrt{a^{2}k_{y}^{2}+b^{2}k_{x}^{2}}\\
ak_{y}+ibk_{x}
\end{array}\right].
\end{eqnarray}
The pseudo spin texture are calculated as
\begin{equation}
\langle\bm{\sigma}\rangle_{n}\propto n(ak_{y},\;bk_{x},\;0).
\end{equation}
The relation between real spin and pseudo spin are $\langle s_{x},s_{y},s_{z}\rangle=\langle g_{xx}\sigma_{x},g_{yy}\sigma_{y},g_{zz}\sigma_{z}\rangle$, 
with $g_{xx}=0.0687$, $g_{yy}=-0.1223$, $g_{zz}=-0.1484$. 
Then the real spin texture are calculated as
\begin{equation}
\langle{\bm{s}}\rangle_{\bm{k},n}\propto n(ag_{xx}k_{y}\;bg_{yy}k_{x},\;0),\label{eq:-5-2}
\end{equation}
as shown in Fig.~\ref{fig:SmB6_SST} with 
parameters 
$a=-0.04$,
$b=0.05$.

The nonzero matrix elements in Eq.~(\ref{eq:final_states}) has the
following relations under $C_{2v}$ symmetry constraint
\begin{align}
_{f}\langle+\frac{1}{2}|\mathcal{P}_{+}|+\frac{3}{2}\rangle_{i} & ={}_{f}\langle-\frac{1}{2}|\mathcal{P}_{-}|-\frac{3}{2}\rangle_{i}=p,\label{eq:-2}\\
_{f}\langle+\frac{1}{2}|\mathcal{P}_{-}|+\frac{3}{2}\rangle_{i} & ={}_{f}\langle-\frac{1}{2}|\mathcal{P}_{+}|-\frac{3}{2}\rangle_{i}=q.
\end{align}
Then the final states in Eq.~(\ref{eq:final_states}) can be written
as
\begin{eqnarray}
|f\rangle & = & \frac{1}{\sqrt{2(a^{2}k_{y}^{2}+b^{2}k_{x}^{2})}}\left[\begin{array}{c}
n\sqrt{a^{2}k_{y}^{2}+b^{2}k_{x}^{2}}\big(q\mathcal{A}_{+}+p\mathcal{A}_{-}\big)\\
(ak_{y}+ibk_{x})\big(p\mathcal{A}_{+}+q\mathcal{A}_{-}\big)
\end{array}\right].\label{eq:-14-1-1}
\end{eqnarray}
With this formula, the spin polarization of photon-emitted electron
can be easily calculated as discussed before.

\begin{figure}[H]
\includegraphics[width=1\columnwidth]{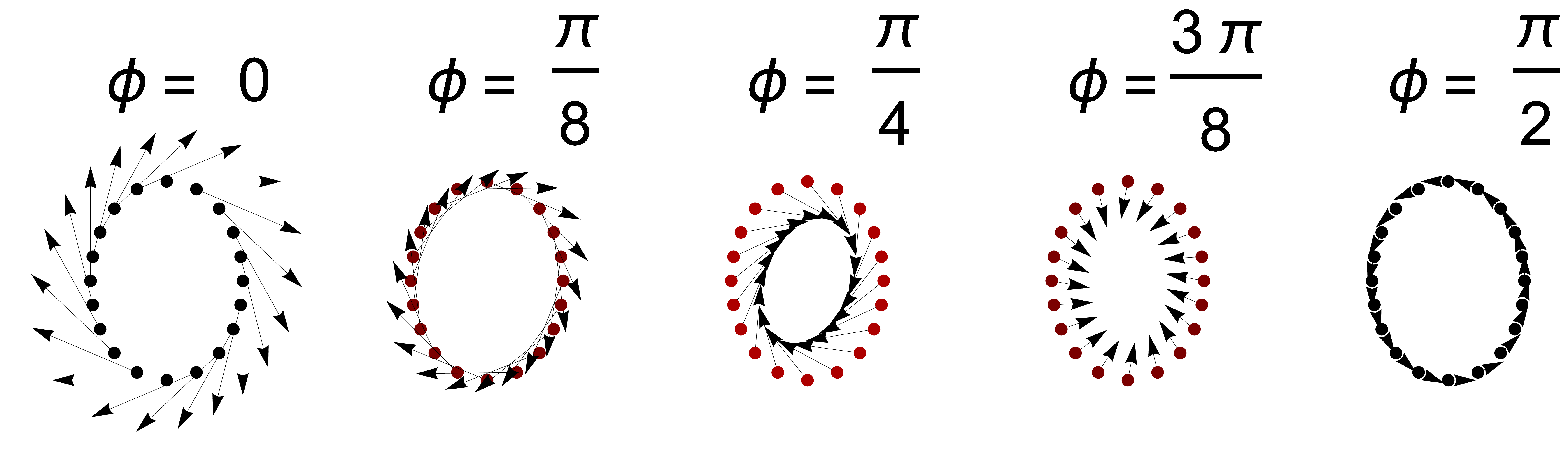} \protect\caption{\label{fig:SmB6_X_s.pol}(Color online) Spin polarization of photon-electrons with
$\sigma$-polarized light. Arrows indicate the spin component in the $xy$ plane.
Color indicate the spin $z$ component: red for $\langle \tau_{z}\rangle<0$,
blue for $\langle \tau_{z}\rangle>0$ and black for $\langle \tau_{z}\rangle=0$. }
\end{figure}

(i) For $\sigma$-polarized light, the spin vector of the photo-emitted
electrons is given as
\begin{eqnarray}
\langle\bm{\tau}_{x}\rangle_{f} & = & n\bigg[ak_{y}\big((p^{*}q+pq^{*})-cos2\phi(|p|^{2}+|q|^{2})\big)\nonumber \\
 &  & +bk_{x}sin2\phi(|p|^{2}-|q|^{2})\bigg]/\sqrt{a^{2}k_{y}^{2}+b^{2}k_{x}^{2}},\\
\langle\bm{\tau}_{y}\rangle_{f} & = & n\bigg[bk_{x}\big((p^{*}q+pq^{*})-cos2\phi(|p|^{2}+|q|^{2})\big)\nonumber \\
 &  & +ak_{xy}sin2\phi(|q|^{2}-|p|^{2})\bigg]/\sqrt{a^{2}k_{y}^{2}+b^{2}k_{x}^{2}},\\
\langle\bm{\tau}_{z}\rangle_{f} & = & -i(p^{*}q-pq^{*})sin2\phi.\label{eq:-52}
\end{eqnarray}
The spin texture is shown in Fig. \ref{fig:SmB6_X_s.pol} with 
parameters 
$p=0.16 - 0.1i$ and
$q=-0.5 + 0.1i$ with unit $\rm A\cdot m$.

\begin{figure}[H]
\centering\includegraphics[width=1\columnwidth]{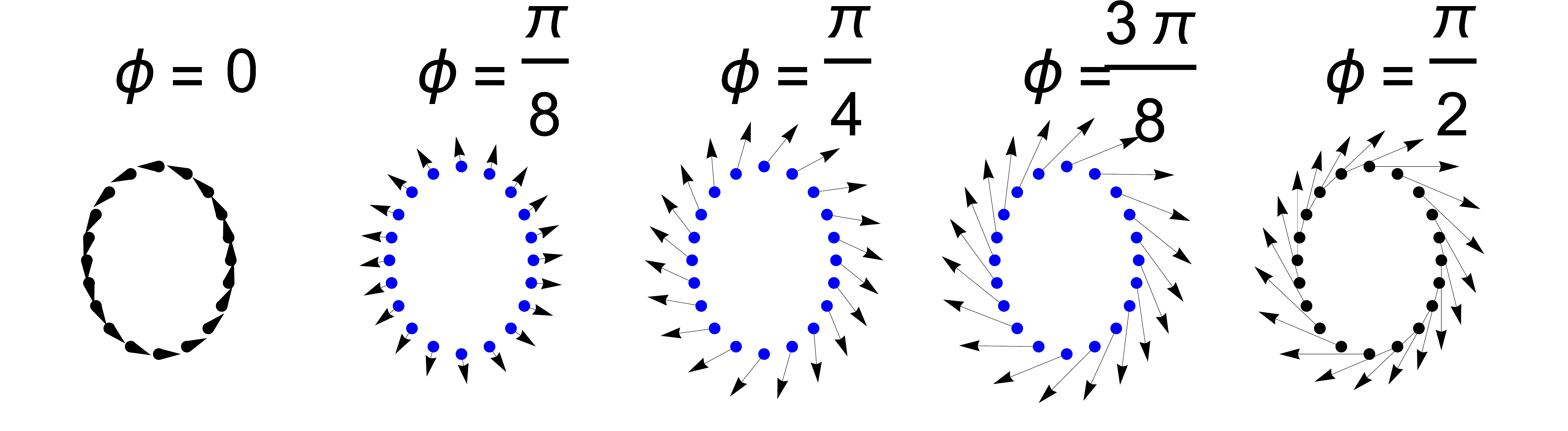}
\protect\caption{\label{fig:SmB6_X_p.pol}(Color online) Spin polarization of photon-electrons with
$\pi$-polarized light. Arrows indicate the spin component in the $xy$ plane.
Color indicate the spin $z$ component: 
red for $\langle \tau_{z}\rangle<0$,
blue for $\langle \tau_{z}\rangle>0$ and 
black for $\langle \tau_{z}\rangle=0$.}
\end{figure}

(ii) For $\pi$-polarized light, the spin vectors are
\begin{eqnarray}
\langle\bm{\tau}_{x}\rangle_{f} & = & n\;cos^{2}\theta\bigg[ak_{y}\big(p^{*}q+pq^{*}+cos2\phi(|p|^{2}+|q|^{2})\big)\nonumber \\
 &  & -bk_{x}sin2\phi\big(|p|^{2}-|q|^{2}\big)\bigg]/\sqrt{a^{2}k_{y}^{2}+b^{2}k_{x}^{2}},\\
\langle\bm{\tau}_{y}\rangle_{f} & = & n\;cos^{2}\theta\bigg[bk_{x}\big(p^{*}q+pq^{*}+cos2\phi(|p|^{2}+|q|^{2})\big)\nonumber \\
 &  & -ak_{y}sin2\phi\big(|q|^{2}-|p|^{2}\big)\bigg]/\sqrt{a^{2}k_{y}^{2}+b^{2}k_{x}^{2}},\\
\langle\bm{\tau}_{z}\rangle_{f} & = & icos^{2}\theta(p^{*}q-pq^{*})sin2\phi.\label{eq:-51}
\end{eqnarray}
The spin texture is shown in Fig. \ref{fig:SmB6_X_p.pol}. 

%
(iii) For the RCP/LCP light, 
the nonzero CD spectrum can be obtained by considering the 
high order perturbations in the wave function as shown in Eq.~(\ref{eq:basis_1k}). Keeping the nonzero matrix elements listed below
\[
\langle+\frac{1}{2}|\mathcal{P}_{+}|-\frac{1}{2}\rangle=\langle-\frac{1}{2}|\mathcal{P}_{-}|+\frac{1}{2}\rangle=d_{1}
,\]
\[
\langle+\frac{1}{2}|\mathcal{P}_{+}|+\frac{3}{2}\rangle=\langle-\frac{1}{2}|\mathcal{P}_{-}|-\frac{3}{2}\rangle=d_{2}
,\]
\[
\langle+\frac{1}{2}|\mathcal{P}_{+}|-\frac{5}{2}\rangle=\langle-\frac{1}{2}|\mathcal{P}_{-}|+\frac{5}{2}\rangle=d_{3}
,\]
\[
\langle+\frac{1}{2}|\mathcal{P}_{-}|-\frac{1}{2}\rangle=\langle-\frac{1}{2}|\mathcal{P}_{+}|+\frac{1}{2}\rangle=d_{4}
,\]
\[
\langle+\frac{1}{2}|\mathcal{P}_{-}|+\frac{3}{2}\rangle=\langle-\frac{1}{2}|\mathcal{P}_{+}|-\frac{3}{2}\rangle=d_{5}
,\]
\[
\langle+\frac{1}{2}|\mathcal{P}_{-}|-\frac{5}{2}\rangle=\langle-\frac{1}{2}|\mathcal{P}_{+}|+\frac{5}{2}\rangle=d_{6}
,\]
\[
\langle+\frac{1}{2}|\mathcal{P}_{z}|+\frac{1}{2}\rangle=\langle-\frac{1}{2}|\mathcal{P}_{z}|-\frac{1}{2}\rangle=d_{7}
,\]
\[
\langle+\frac{1}{2}|\mathcal{P}_{z}|-\frac{3}{2}\rangle=\langle-\frac{1}{2}|\mathcal{P}_{z}|+\frac{3}{2}\rangle=d_{8}
,\]
\begin{equation}
\langle+\frac{1}{2}|\mathcal{P}_{z}|+\frac{5}{2}\rangle=\langle-\frac{1}{2}|\mathcal{P}_{z}|-\frac{5}{2}\rangle=d_{9},
\end{equation}
the CD values can be obtained by using Eq.~(\ref{eq:final_states}), Eq.~(\ref{eq:I_eta}) and Eq.~(\ref{eq:I_CD_bise}). 
The expressions for the CD values around $\bar{Y}$ and $\bar{X}$
point are very lengthy and will not be given here. We only show the numerical results in Fig.~\ref{fig:SmB6_GM_CD} with parameters
$a_1=-0.2$,
$a_2=0.7$, 
$a_3=0.42$,  
$d_1=(-0.1+0.1i)$,
$d_5=(-0.5+0.1i)$,
$d_6=(-0.02+0.04i)$,
$d_7=(0.2-0.3i)$,
$d_2=(0.16-0.1i)$,
$d_3=(0.06-0.02i)$,
$d_4=(0.1-0.05i)$,
$d_8=(0.06-0.02i)$ and
$d_9=(0.01-0.03i)$ where $d_i$ with unit $\rm A\cdot m$. 
These parameters well reproduce the experimental result in Ref.~\onlinecite{FengDL_SmB6_CD_ARPES_2013_NatComm}.

\section{conclusion}

To summarize, 
we discussed three different definition of the spin texture for the topological surface states, namely the pseudo spin and real spin orientation 
for electronic states inside the crystal and that of the photo-emitted 
electrons in the vacuum.
Taking \bise and \smb as examples, we revealed that the above three spin 
textures are different and should be clarified rigorously and studied 
separately.
By considering the symmetry properties of the photo-emission matrix element, 
we calculated the spin polarization and CD spectrum of the photo-electrons 
for these two compounds 
which can be observed in the spin-resolved and CD ARPES experiment.

\vspace{3mm}
\noindent\textit{Acknowledgments ---}
This work was supported by  the National Natural Science Foundation
of China, the 973 program of China 
(No.2013CB921700), and the ``Strategic Priority Research Program (B)" of
the Chinese Academy of Sciences (No.XDB07020100).
R.Y. acknowledges funding form the Fundamental Research Funds for 
the Central Universities (Grant No.AUGA5710059415).

\appendix

\section{\textup{Some discussions about vector}\textit{ $\bf{A}$}}

As shown in Fig.~\ref{fig:setup} the local coordinate system $x'y'z'$
and the global coordinate system $xyz$ are related by
\begin{eqnarray}
x' & = & (cos\theta cos\phi,\;cos\theta sin\phi,\;-sin\theta),\label{eq:-45}\\
y' & = & (-sin\phi,\;cos\phi,\;0)\label{eq:-46},\\
z' & = & (sin\theta cos\phi,\;sin\theta sin\phi,\;cos\theta).\label{eq:-27}
\end{eqnarray}
For $\pi$-polarized light, the vector $\bm{A}$ are expressed as\textbf{
}
\begin{equation}
\bm{A}_{\pi}^{\prime}=(A_{0}cos\omega t,\;0,\;0)\overset{F.T.}{\Longrightarrow}(A_{0},\;0,\;0),\label{eq:-28}
\end{equation}
in the local coordinate system and
\begin{equation}
\bm{A_{\pi}}=A_{0}(cos\theta cos\phi,\;cos\theta sin\phi,\;-sin\theta).\label{eq:A-pi}
\end{equation}
in the global coordinate system.

For $\sigma$-polarized light, we have
\begin{equation}
\bm{A}_{\sigma}^{\prime}=(0,\;A_{0}cos\omega t,\;0)\overset{F.T.}{\Longrightarrow}(0,\;A_{0},\;0),\label{eq:-30-1}
\end{equation}
in the local coordinate system and
\begin{equation}
\bm{A}_{\sigma}=A_{0}(-sin\phi,\;cos\phi,\;0),\label{eq:-8}
\end{equation}
in the global coordinate system.

For left and right circular polarized light we get
\begin{equation}
\bm{A}_{\eta}^{\prime}=A_{0}(cos\omega t,\;\eta\;sin\omega t,\;0)\overset{F.T.}{\Longrightarrow}A_{0}(1,\;-\eta i,\;0),\label{eq:-32}
\end{equation}
in the local coordinate system, where $\eta=\pm1$ indicate the right
and left circular polarized light. In the global coordinate system
$\bm{A}$ has the following form:
\begin{align}
\bm{A}_{\eta}=  A_{0}(cos\theta cos\phi+\eta\;isin\phi,
  \;cos\theta sin\phi-\eta\;icos\phi,\;-sin\theta).
\end{align}


\bibliographystyle{apsrev4-1}
\bibliography{refs}

\end{document}